\documentclass[pra,
a4paper,
twocolumn,
superscriptaddress]{revtex4-1}
\usepackage{amssymb}
\usepackage{amsmath}
\usepackage{amsfonts}
\usepackage{braket}
\usepackage{graphicx}
\usepackage{bm}
\usepackage{color}
\usepackage{multirow}
\usepackage{natbib}

\usepackage[normalem]{ulem}
\usepackage{verbatim}
\def \be {\begin{equation}}
\def \ee {\end{equation}}

\def \p {\partial}
\def \BEA {\begin{eqnarray}}
\def \EEA {\end{eqnarray}}

\def \BC {\begin{cases}}
\def \EC {\end{cases}}

\newcommand{\aleq}[1]{
\begin{equation}
    \begin{aligned}
    #1
    \end{aligned}
\end{equation}
}

\usepackage{xr-hyper}    

\makeatletter
\newcommand*{\addFileDependency}[1]{
  \typeout{(#1)}
  \@addtofilelist{#1}
  \IfFileExists{#1}{}{\typeout{No file #1.}}
}
\makeatother

\newcommand*{\myexternaldocument}[1]{%
    \externaldocument{#1}%
    \addFileDependency{#1.tex}%
    \addFileDependency{#1.aux}%
}

\myexternaldocument{SuppMat} 

\usepackage{hyperref}

\begin{document}
\title
{Coherent spin transport through helical edge states of topological insulator}

\author{R.\,A.~Niyazov $^{\ast}$\,}
\address{Department of Physics, St. Petersburg State University, St. Petersburg 198504, Russia}
\affiliation{Ioffe Institute,
194021 St.~Petersburg, Russia}
\affiliation{NRC ``Kurchatov Institute", Petersburg Nuclear Physics Institute, Gatchina
188300, Russia}
\author{D.\,N.~Aristov}
\affiliation{Ioffe Institute,
194021 St.~Petersburg, Russia}
\affiliation{NRC ``Kurchatov Institute", Petersburg Nuclear Physics Institute, Gatchina
188300, Russia}
\address{Department of Physics, St. Petersburg State University, St. Petersburg 198504, Russia}
\author{V.\,Yu.~Kachorovskii }
\affiliation{Ioffe Institute,
194021 St.~Petersburg, Russia}
\affiliation{CENTERA Laboratories, Institute of High Pressure Physics, Polish Academy of Sciences, 01-142 Warsaw, Poland
 }


\begin{abstract}
We study coherent spin transport  through helical edge  states of topological insulator tunnel-coupled to metallic leads.
 We demonstrate that unpolarized incoming  electron beam  acquires  finite polarization after  transmission through  such a setup  provided that  edges  contain at least one   magnetic impurity.    The finite polarization  appears even in the fully classical regime and is therefore robust to dephasing.  There is also a quantum
 magnetic field-tunable
 contribution to the polarization, which
 shows sharp identical   Aharonov-Bohm  resonances as a function of  magnetic flux---with the period $hc/2e$---and survives at relatively high temperature.  We demonstrate that this tunneling interferometer can be described in terms of ensemble of flux-tunable   qubits giving equal contributions to conductance and spin polarization.  The number of active qubits participating in the charge and spin transport is given by the ratio of the temperature and the level spacing.    The interferometer can  effectively operate at high temperature and can be used for quantum calculations.  In particular,  the ensemble of  qubits  can be described by a single Hadamard operator. The obtained results open wide avenue for applications in the area of quantum computing.
 \end{abstract}

 \maketitle

\section{Introduction}

Quantum information processing attracts enormous interest of a broad scientific community \cite{Grumbling2019}.
Although the promise of quantum computers was recognized about thirty years ago, the real breakthrough in creation of  their key elements --- networks of  coherent  spin qubits ---  was achieved only in the last decade~\cite{Zwanenburg2013}. The principal obstacle for further progress is connected  with  fast spin relaxation and dephasing, which prevent  creation of spin polarization and coherent spin transmission over long distances.  Another challenging  yet unsolved task of primary importance for  information processing and quantum networking is all-electrical control of the electron spins \cite{N-1,N1,N0}.

 An effective low-cost room-temperature solution of these   problems   would    allow for tunable  coherent transmission of the spin polarization over long distances.
Ever since the proposal of spin field effect transistor (SpinFET) \cite{L1},
numerous attempts to achieve  coherent spin transmission and all-electrical manipulation by using  setups of various design were unsuccessful  \cite{L4,L5,L6,L7,L8,L10,L11}. In semiconductor devices spin polarization  usually originates from spin-orbit coupling and is never sufficiently large, in particular due to low efficiency of the spin injection \cite{L9}. It can be somewhat increased by using non-electrical elements such as ferromagnetic contacts, which however dramatically deteriorate transport properties of the system. Furthermore,
injected  polarization rapidly decays  due to  spin relaxation processes.

In this Letter, we propose  essential steps towards solving 
several critical problems of  quantum information processing: spin filtering,  long-distance spin transfer, and effective spin manipulation. Physically,  spin filter  blocks transmission of particles   with one  spin orientation, say spin-down,    so that  outgoing current  acquires  spin-up polarization.
We introduce a method  for creation of spin-polarized electron beams based on using of helical edge states (HES) of two-dimensional (2D) topological insulator.  Spin transport in HES was already discussed at zero temperature (see Refs.~\cite{An2012a,An2012,Michetti2011,Battilomo2018,Zare2019} and references therein). Here,
we  demonstrate that, remarkably,
the finite spin polarization arises at high temperature, even in the fully classical regime and is therefore robust to dephasing.

The suggested method allows for 100\% spin polarization and therefore  has essential advantages over the
existing approaches to   spin filtering and spin transfer based on  resonant tunneling diodes \cite{N6,N8}, quantum dots \cite{N2,N3}, Y junctions \cite{N4,N38}, and Aharonov-Bohm (AB)  interferometers based on conventional materials \cite{Shmakov2012}. In all these structures   spin polarization   achieved so far
was sufficiently small.
More promising candidates for  spin filtering are
the  quantum point contacts  (QPC) with strong SO interaction  and engineered structures incorporating  QPC  as building blocks \cite{N5,N9,N11,N13,N15,N20}.  Although
 the predicted spin polarization in QPC-based structures operating in   the single-mode
 regime of SpinFET can be quite  high \cite{N13},  one of the main problems in the way of  coherent spin control---fast spin relaxation---remains unresolved. This implies that spin polarization cannot be transferred  over a distance  exceeding the spin relaxation length which is typically not quite large for conventional semiconductors   with SO interaction.

Here, we study spin transport through the edge states of topological insulator and show that  the spin polarization can  be transferred for large distances on the  order of the edge state's length.  This distance can be made  even longer by building arrays of several HES. In contrast to all previous studies of spin-selective transport via HES, we find that large spin polarization can be created and transferred at high temperatures thus opening a wide avenue for 
application in quantum computing. In particular, we demonstrate that obtained results can be formulated in terms of flux-tunable ensemble of qubits giving equal contribution to charge and spin transport.  Our study is a direct generalization of recent research on controlling quantum qubits  by various types of interferometers and using them for   quantum computing
\cite{Foldi2005,Michetti2011,Chen2014,Bautze2014,Baurle2018,Bordone2019,Bellentani2020}.
 In particular, it was predicted that  conventional interferometers with spin-orbit (SO) interaction (or an array  of such interferometers) can be used as one-qubit quantum gates of various types (X-gate, Z-gate, phase gate, and Hadamard gate) \cite{Foldi2005}.  Such  qubits can be controlled by changing the magnetic field and the strength of the SO interaction \cite{Foldi2005,Michetti2011}. Taking into account the  electron-electron interaction  makes it also possible to construct effective two-qubit computational schemes  in   two coupled  interferometers based on conventional materials 
\cite{Bautze2014,Baurle2018},  on edge states of the  integer quantum Hall effect \cite{Bordone2019,Bellentani2020} and on helical states \cite{Chen2014}.   Signatures of electron-electron interaction in HES was already observed  experimentally 
\cite{Stuhler2019,Strunz2020}. 

 The  computational  schemes, proposed so far, imply the control of  so-called flying qubits with a given energy  and can be directly applied at zero temperature.
However, in realistic systems, the  electrons enter the interferometer from thermalized contacts, which implies averaging within the temperature window around the Fermi energy.   Since the phases accumulated by an electron passing through two arms of the interferometer are energy dependent, the question arises whether thermal averaging  violates the efficiency of the proposed computational schemes. This is exactly the question that we address in this work.
We demonstrate that using tunneling interferometers based on helical edge states  allows one 
for transfer of spin polarization  at large distance as well as quantum computing   at high temperatures.    We also find that the energy levels of almost closed interferometer  form an ensemble of  $T/\Delta$ qubits  providing equal contributions into the spin and charge transport.   This means that in HES based setups the interference survives thermal averaging~\cite{commentOrdinary}. 
Hence, using of such interferometers  might  be   a neat way to overcome  the main problems of spin-networking, namely, sensitivity of spin polarization to dephasing and relaxation processes and the requirement of very low temperature.

\section{Results}
 \subsection{Key idea}

We  propose to explore unique properties of HES existing at the edges of 2D topological insulators,
which are materials insulating in the bulk, but  exhibiting  conducting channels at the surface or at the boundaries. In particular,  the 2D  topological insulator phase was predicted in HgTe quantum wells \cite{Kane2005,predicted} and confirmed by direct   measurements of conductance of the  edge states \cite{confirmed} and  by  the experimental analysis of the non-local transport   \cite{Roth2009,Gusev2011,Brune2012,Kononov2015}.
These states  are
 one-dimensional helical channels where the electron spin projection is connected with its velocity, e.g. electrons traveling in one direction are characterized by spin ``up'', while electrons moving in the opposite direction are characterized by spin ``down''.
 Remarkably, the electron transport  via HES is ideal, in the sense that electrons do not experience backscattering from conventional non-magnetic  impurities, similarly to what occurs in edge states of Quantum Hall Effect systems, but without invoking high magnetic fields (for detailed discussion  of properties of  HES see Refs.~\cite{Hasan2010,Qi2011}).

 Hence, in the absence of magnetic disorder, the boundary states are  ballistic and  topologically protected from external perturbations. Due to this key advantage a spin  traveling along the edge does not relax, so that such states  perfectly match the purposes of  quantum spin networking.
Importantly, even a non-magnetic lead splits the incoming electron beam into two parts: right-moving electrons with spin up and left-moving electrons with spin down. If the transmission over one of the shoulders of the system is blocked, say, by inserting a strong magnetic impurity into the upper shoulder, then only the down shoulder remains active and  the spin polarization of outgoing electrons can achieve 100\%.  Remarkably, this mechanism is robust to dephasing and, therefore, works at high temperatures.
We find a quantum contribution to polarization, which shows Aharonov-Bohm oscillations with the magnetic flux piercing the area encompassed by HES and is therefore  tunable by  external magnetic field. This contribution survives at relatively high temperature.

We  also demonstrate that  tunneling interferometer  can be described in terms of ensemble of flux-tunable qubits giving equal contributions to conductance and spin polarization.  The number of active qubits participating in the charge and spin transport is given by the ratio of the temperature and the level spacing.  The interferometer can effectively operate at high temperature and can be used for quantum calculations.  In particular, the ensemble of qubits can be described by a single flux-tunable Hadamard operator. Measurement of the conductance and the spin polarization is one of the ways to read out information about qubit states. 

\subsection{Model}

The   Hamiltonian  of the edge is  given by  $H= \int dx  \left({\cal H}_0+{\cal H}_{\rm imp}\right)$ with 
coordinate $x$ running along  the edge. Here,
\begin{equation}
{\cal H}_0=-iv_\text{F} \left(   \psi^\dagger_{\uparrow}  \partial_x \psi_{\uparrow}-
\psi^\dagger_{\downarrow}  \partial_x \psi_{\downarrow} \right),
\end{equation}
is the unperturbed  HES Hamiltonian 
with   the  Fermi velocity  $v_{\rm F}$. 
For simplicity, we assume that interferometer contains     classical impurities  with 
  large magnetic moments $\mathbf M_n,$  
  $|\mathbf M_n|=M\gg 1$ (a small ferromagnetic island can serve as such an impurity),
     neglecting  feedback  effect related to the dynamics of this moment caused by  exchange  interaction with the ensemble of  right- and left-moving  electrons (for infinite HES  this  effect was discussed in Ref.~\cite{kur3}). 
Then, the isotropic exchange interaction with magnetic impurities located at points  $x_n$ has the form
\begin{equation}
    {\cal H}_{\rm imp}=  g \sum \limits_n \boldsymbol{\sigma} \mathbf{M_n} \, \delta (x-x_n),
\label{ham}
\end{equation}
where $g$ is the coupling constant and $\mathbf M_n= M(\sin \eta_n \cos \varphi_n, \sin \eta_n \sin \varphi_n, \cos \eta_n ). $ Here angles $\eta_n$  and  $\varphi_n$  describe direction of $\mathbf M_n.$

\begin{figure}
\includegraphics[width=0.9\columnwidth]{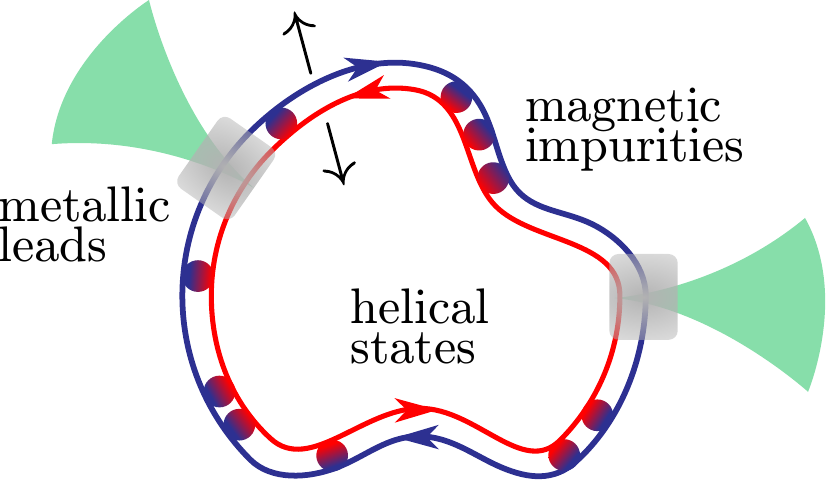}
\caption{\label{fig:densityring}
(Color online)
Helical edges states tunnel-coupled to the metallic point contacts. The   magnetic impurities  are marked by  dots. }
\end{figure}

 In the general case,  the edge contains randomly distributed magnetic impurities shown by  dots in Fig.~\ref{fig:densityring}. However,  as we demonstrate below, the simplest case of an interferometer containing a single impurity captures basic physics of the problem.  At the same time, this case  is the most realistic, since we discuss non-magnetic materials. 
 Hence, we start with discussion of the   interferometer with the single impurity placed in the upper shoulder.    By using  Eq.~\eqref{ham},  one can find the 
  scattering matrix of this impurity \cite{comment1}
\be
\hat S_M= \begin{pmatrix}
  e^{i \alpha} \cos \theta &  i \sin \theta ~e^{-i\varphi}\\
i \sin \theta ~e^{i\varphi} & e^{-i \alpha} \cos \theta\\
\end{pmatrix},
\label{SM}
\ee
  where  $\alpha$ is the forward scattering phase  and $\sin^2 \theta$ is the
  backscattering  probability. For weak impurity with $\rho_0= g M/v_F \ll 1,$  one gets:  $\alpha \approx \rho_0 \cos \eta \ll 1 $ and 
  $\theta \approx \rho_0 \sin \eta \ll 1.$   
  
\begin{figure}
\includegraphics[width=0.8\columnwidth]{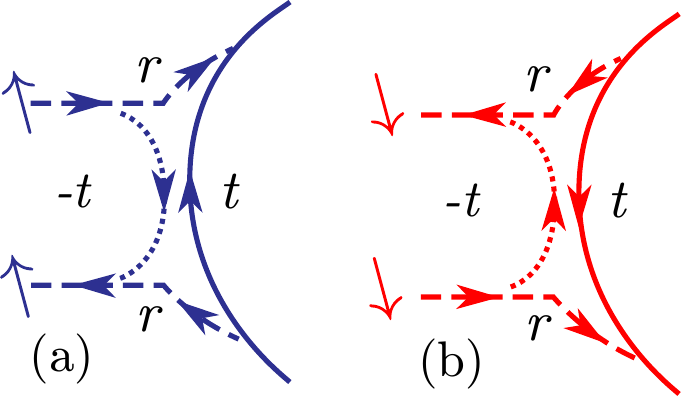}
\caption{\label{fig:pointcontact}
(Color online)
Non-magnetic point contact between the helical ring and the spinful wire. 
Different spin projections  (shown by blue and red color)  do not mix at the contact and correspond to  electrons propagating  along the helical edge in the opposite directions.      The  spin-independent  
amplitudes $t$ and $r$ obey $t^2+r^2=1$. }
\end{figure}

The spin transport through  HES   of a 2D topological insulator assumes   tunnel coupling
 to leads (see Fig.~\ref{fig:densityring}).    The tunneling  conductance of this setup
 is given by
$G=2 \times  ({e^2}/{h})  \mathcal T,$
 where factor $2$ corresponds to two conducting channels. For the case of  spin-unpolarized contacts, the
transmission coefficient, $\mathcal T$, can be represented as
  an average over incoming spin polarizations
  $\mathcal T= ({\mathcal T_\uparrow + \mathcal T_\downarrow})/2  .$
 Here
 $ \mathcal T_\beta= \sum _\alpha T_{\alpha \beta}  =\sum _\alpha\langle|t_{\alpha\beta} (\epsilon) |^2 \rangle_\epsilon,$
 $\alpha,\beta=\uparrow,\downarrow,$  $\langle \cdots \rangle_\epsilon= -\int d\epsilon (\cdots) \p_\epsilon f_F(\epsilon)$,    $f_F(\epsilon)$ is the Fermi   function and $t_{\alpha\beta}$ is a  spin-dependent transition amplitude.
 The spin polarization of outgoing  electrons reads \cite{commentP}
  \be
P_z=({\mathcal T_{ \uparrow}-\mathcal T_{ \downarrow}})/({\mathcal T_{ \uparrow}+\mathcal T_{ \downarrow}}),
\label{polarization}
\ee
  where $z-$axis coincides with direction of spin at the position of outgoing contact.
    We consider  nonmagnetic  leads,
  thus assuming that  different spin projections do not mix at the tunneling contacts, so that   electrons entering the edge with opposite spins move in the opposite directions (see Fig.~\ref{fig:pointcontact}). 
  Such contacts are characterized by spin-independent  amplitudes $r$ and $t,$  obeying $|t|^2+|r|^2=1.$
We assume that $t$  and  $r$ are real   and   positive and
   parameterize them as follows \cite{comment2}:
  $r=\sqrt{1-e^{-2\lambda}}, \quad
 t=e^{-\lambda},
  \quad 0< \lambda<\infty.$

 We will study both classical and quantum contributions to the spin polarization. The quantum contribution  is sensitive to magnetic field due to the AB effect. Hence our setup represents an example of AB interferometer built on HES. 
The  form and shape  of the AB oscillations  strongly depend on the relation between temperature $T$  and level spacing    $\Delta=2\pi v_F/L,$  which is controlled by total interferometer circumference $L$
  and  the Fermi velocity  $v_F.$   Let us do some estimates.  For  $L =10~ \mu$m  and $v_F=10^7$ cm/s, we get $\Delta \approx 3 $ K. As seen from this estimate, the case
  \be T\gg \Delta  \label{TggD}\ee
  is  much more interesting for   possible applications. We will focus on this case throughout the article.   
There is also upper limitation for  temperature. For good quantization,  $T$  should be much smaller than the bulk gap of the topological insulator: $T\ll \Delta_{\rm b}$.    For the first time quantum spin Hall effect was observed in structures based on HgTe/CdTe~\cite{Konig2007} and InAs/GaSb~\cite{Knez2011}, which had a rather narrow bulk gap, less than 100 K. Substantially large values were observed recently in WTe$_2,$ where gap of the order of 500 K was observed~\cite{Wu2018},  and in bismuthene grown on a SiC (0001) substrate, where a bulk gap of about 0.8 eV was demonstrated~\cite{Reis2017,Li2018} (see also recent discussion in Ref.~\cite{Stuhler2019}).    Thus, recent experimental studies unambiguously indicate the possibility of transport through HES at room temperature, when the condition $\Delta_{\rm b} \gg T\gg \Delta$,  needed for applicability of our theory, can be easily satisfied.  Importantly,  this condition ensures the universality of spin and charge transport (see 
discussion in Ref.~\cite{my-cond}), which do not depend on details of the systems, in particular, on the device geometry.

\subsection{Tunneling conductance}

Recently, we discussed dependence of the tunneling conductance $G$ of such a setup on the     external   magnetic
flux $\Phi$ piercing the area encompassed by  edge states
    \cite{my-cond}. For consistency, we briefly summarize   main results  of Ref.~\cite{my-cond} here.
  We have demonstrated
 the existence of interference-induced effects, which are robust to the  temperature, i.e. survive under the condition Eq.~\eqref{TggD},
  and can therefore be obtained for relaxed experimental conditions (for discussion of this regime in conventional interferometers see  Refs.~\cite{jagla,dmitriev,Shmakov2013,Dmitriev2015,SCS2017}).
Specifically, we  have  found that   $G$ is  structureless in   ballistic case but shows periodic dependence on dimensionless flux  $\phi=\Phi/\Phi_0$ (here,   $\Phi_0=hc/e$ is the flux quantum), with the period $1/2,$
 in the presence of
   a single    magnetic impurity in one of the interferometer's shoulders.   Such a weak impurity  can be taken into account perturbatively provided that  $\theta \ll \rm{max}(\lambda,1).$   The resulting   analytical expression for the transmission coefficient reads \cite{my-cond}
 \be
 \mathcal T=\tanh \lambda - \frac{ {\tilde \theta}^2}{2} \tanh^2\lambda
 \label{T-pert}
 \ee
where
${\tilde \theta}^2 = { \theta}^2 (1+\mathcal C|_{\theta=0})$
and
\be
{\cal C}=\frac{t^4 e^{4 i\pi\phi}}{1- t^4\cos^2\theta e^{4i\pi\phi} }+\frac{t^4 e^{-4i\pi\phi}}{1- t^4\cos^2\theta e^{-4i\pi\phi} },
\label{cooperon}
\ee
represents ``ballistic Cooperon'' \cite{my-cond} which
is the  interference contribution of the processes  in which the  electron wave splits  at the impurity into two parts passing the setup in the opposite directions and returning to impurity after  a number of  revolutions with equal winding  numbers (see Fig.~6 of Ref.~\cite{my-cond}).
 The factor
 \be
\frac{{\tilde \theta}^2}{  { \theta}^2}=1+{\cal C}|_{\theta=0}=
\frac{\sinh (4\lambda)}{\cosh (4\lambda) -\cos(4\pi\phi)},
\label{factor}
\ee
describes coherent  enhancement of backscattering probability caused by multiple  returns to the impurity.
This enhancement has a purely quantum nature.  The classical limit, when all interference processes are neglected, can be obtained by averaging $ \mathcal T $ over flux.  Having in mind that $\langle  \mathcal C \rangle_\phi =0,$
we find that "classical" conductance is given by Eq.~\eqref{T-pert} with the replacement ${\tilde \theta}^2 \to \theta^2.$
   Hence, in the  perturbative regime, $\mathcal T$   obeys $1/2-$flux periodicity  $\mathcal T(\phi+1/2)= \mathcal T(\phi)$ and
  shows sharp  identical  antiresonances at integer and half-integer values of $\phi$ in the limit of weak tunneling coupling, $\lambda \ll 1.$
  In the latter limit,
  the non-perturbative effects lead  to appearance of the   additional contribution  $2\theta^2.$
in the denominator of  Eq.~\eqref{factor} \cite{my-cond}.  Physically, this corresponds to the broadening  of  the antiresonances because of multiple coherent scattering events.

\subsection{Spin polarization}
Next, we discuss the spin polarization of outgoing electrons.
We will limit ourselves with discussion of non-interacting electrons     focusing  on   high temperature case. For discussion of spin polarization in the   low temperatures case see  Refs.~\cite{Chu2009, Masuda2012,Dutta2016,Bjoernson2018,Zhou2019}, while interaction-induced \cite{Ronetti2016}  and quantum pumping generated \cite{Ronetti2017}  spin currents were considered for  Fabry-P\'erot
geometry at $\phi=0.$  We will demonstrate  that the  finite polarization  appears even in the fully classical regime and therefore robust to dephasing.  There also exists quantum contribution to polarization which  survives at relatively large temperature    and        is tunable by magnetic flux piercing the interferometer. Specifically, we will demonstrate  that similar to tunneling conductance the  quantum contribution to the polarization shows sharp identical resonances as a function of    magnetic flux with maxima (in the absolute value)  at integer and half-integer values of the flux.

\begin{figure}[ht]
\centerline{\includegraphics[width=0.9\columnwidth]{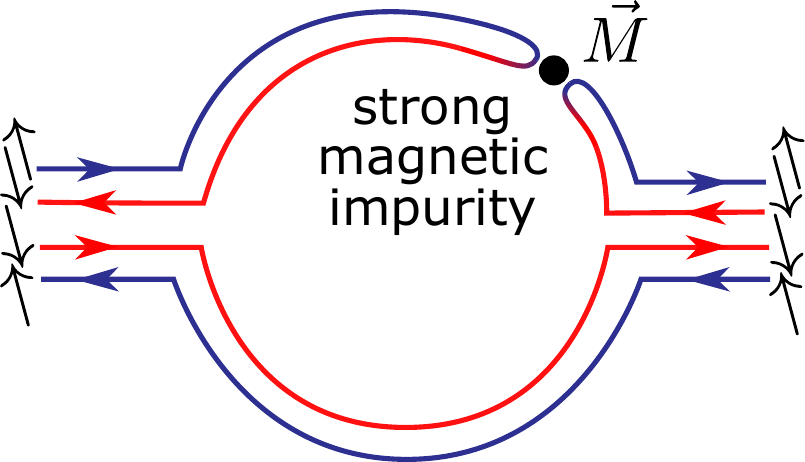}}
\caption{
(Color online)
Strong magnetic  impurity blocks transmission of one component of the electron spin. For open setup, $\lambda =\infty,$ this leads to 100\% polarization. Polarization reverses sign, when strong impurity is moved from upper to lower shoulder.    }
\vspace*{0.3cm}
\label{fig-strong-impurity}
\end{figure}

In order to illustrate our approach, we consider a  single  impurity placed in the upper  shoulder of the interferometer and discuss a simple limiting case: $\lambda=\infty,~\theta=\pi/2$ (strong impurity, open interferometer). In this case, $t=0$ and $r=1,$ so that electrons with spin up (down) can go only through   upper (lower) shoulder of interferometer (see Fig.~\ref{fig:pointcontact}). On the other hand     probability  of backscattering by the impurity is given by $\sin^2\theta=1,$ so  that impurity fully blocks transmission through the upper shoulder (see Fig.~\ref{fig-strong-impurity}). Hence,  such a setup serves as ideal spin filter:   the transmission of electrons  with spin up is blocked while spin-down electrons can freely pass through the interferometer.  Consequently, the  outgoing    polarization reaches 100\%. Evidently, this is a classical result which is not sensitive to dephasing. At the same time, fully polarized electron beam corresponds to a pure quantum spin state. In other words, even in the classical regime, the  interferometer can create pure quantum states within the discussed limiting case.   
Below, we present  detailed calculations  of the spin polarization for a number  of other  cases.

Results of Ref.~\cite{my-cond} can be easily generalized for calculation of spin polarization. For a weak impurity placed in the upper shoulder of interferometer, direct summation of amplitudes in a full analogy with Ref.~\cite{my-cond}   yields  in the lowest  order in $\theta^2$:
\be
T_{\alpha \beta}=\delta_{\alpha\beta}\tanh \lambda  -   \frac{ \alpha \beta\exp [\lambda (\alpha+\beta)]}{4 \cosh^2\lambda}~{\tilde \theta}^2
\label{|tab|^2}
\ee
where $\alpha,\beta =\pm 1,$ for spin up and down, respectively.
Classical  probabilities  $ \langle T_{\alpha\beta} \rangle_\phi $ are given by
Eq.~\eqref{|tab|^2}
with the replacement ${\tilde \theta}^2 \to \theta^2.$

The  perturbative in $\theta^2$ spin polarization can be found from Eqs.~\eqref{polarization} and \eqref{|tab|^2}:
\aleq{\label{Pz1}
&P_z= -  \frac{{\tilde \theta}^2}{2} =- \frac{\theta^2}{2}\frac{\sinh (4\lambda)}{\cosh (4\lambda) -\cos(4\pi\phi)}.
}
As is seen from this equation, polarization shows  sharp identical antiresonances at  integer and half-integer values of flux  for weak tunneling
coupling, $\lambda \ll 1,$ and weak AB oscillations for almost open setup, $\lambda \gg 1.$
 Analogous calculation for a single impurity  with the same strength, $\theta,$   placed in the lower shoulder of the interferometer yields Eq.~\eqref{Pz1} with the opposite sign.
 In the classical regime, the polarization is simply given by $P_z=\pm \theta^2/2,$ with the sign  determined by the position of impurity.    One can follow the evolution of polarization from quantum to classical case by introducing a  dephasing process with the rate $\Gamma_\varphi$ which suppresses ``ballistic Cooperon''.  Technically, this means replacement $\lambda \to \lambda +\lambda_\varphi$  in Eq.~\eqref{Pz1}, where $\lambda_\varphi= \pi \Gamma_\varphi /2\Delta $ (see Ref.~\cite{my-cond}). For $\lambda_\varphi \to \infty$ we restore the classical result. Away from the resonant points [more precisely, for  $\cos (4\pi \phi) <0$], dephasing leads to the increase of polarization because the interference  for such values of $\phi$ is destructive.

Microscopical calculation of  $\Gamma_\varphi$ in HES is a  non-trivial question. 
     In conventional systems, including  infinite single-channel quantum  wires,   dominates dephasing  caused by electron-electron scattering. In HES, such dephasing is suppressed for the same reason as ordinary impurity backscattering. Nonzero (very slow) dephasing due to  electron-electron interaction arises only when Rashba-type terms are present and slow energy dependence of these terms on energy is  taken into account \cite{Schmidt2012,mirlin1}. Additional suppression of the interaction-induced dephasing is expected due to finite geometry of the setup similar to the case of conventional single-channel interferometers \cite{dmitriev}.   A very slow  dephasing  occurs due to the dynamics of the magnetic impurity. Such dynamics  can arise  due to  the interaction directly with the conduction electrons \cite{kur3} and  due to the presence of a magnetic bath \cite{my-cond}. In the latter case,  assuming that  the averaged magnetic moment of impurity relaxes as $\langle \mathbf M(0) \mathbf M(t)\rangle = M^2 \exp(-\Gamma_0 t)$ one gets $\Gamma_\varphi=\Gamma_0$ \cite{my-cond}. 
     Importantly, all proposed mechanisms lead to dephasing rate significantly slower (at least in the framework of theoretical models) than in conventional systems. 

Let us now consider a setup with a number of randomly distributed impurities. We start our   discussion with the  classical regime ($\lambda_\varphi\to \infty$).
 One  finds then $T_{\alpha\beta}$   as the sum over contributions from classical  trajectories   propagating  clockwise and counterclockwise and experiencing collisions by magnetic impurities  with forward   probability $\cos^2 \theta $   and backward probability  $\sin^2\theta. $
   Relations between classical currents flowing from different sides of the impurity read:
$J_{n+1}^{\uparrow}=\cos^2\theta J_n^{\uparrow}+\sin^2\theta J_{n+1}^{\downarrow},$
~$J_{n}^{\downarrow}=\sin^2\theta J_{n}^{\uparrow}+\cos^2\theta J_{n+1}^{\downarrow}.$
The vectors  $\mathbf J_n =    (J_{n}^{\uparrow}, J_{n}^{\downarrow})$   and $\mathbf J_{n+1} =    (J_{n+1}^{\uparrow}, J_{n+1}^{\downarrow}) $ are thus connected by the classical transfer  matrix
\be
\hat W_{cl} (\theta)= 1+ \tan^2\theta \hat P,\quad \hat P=\left(
                                              \begin{array}{cc}
                                                -1 & 1 \\
                                                -1 & 1 \\
                                              \end{array}
                                            \right).
                                            \label{hat-T}
\ee
It obeys simple multiplication rule,
$\hat W_{cl} (\theta_1)  \hat W_{cl}  (\theta_2 ) =\hat W_{cl} (\theta), $$  \quad \tan^2(\theta)=  \tan^2(\theta_1) + \tan^2(\theta_2).$
 Let us consider   setup containing $N_u$   impurities in the  upper  shoulder, characterized by $\theta_1, \ldots,\theta_{N_u}$
 and $N_l$ in the lower one characterized  by  $\theta_1^\prime, \ldots,\theta_{N_u}^\prime$.
  Due to multiplicativity property
 one can equivalently consider  setup with two impurities having effective  strengths
   $g_u=\sum^{N_u}_{n=1}  \tan^2 (\theta_n)$, and  $g_l=\sum^{N_l}_{n=1}  \tan^2 (\theta^\prime_n),$
  placed respectively  in the upper and lower shoulder of the interferometer. Next, we assume that   current  entering interferometer from the left contact   is unpolarized, and use the  scattering probabilities $r^2$ and $t^2$ to write balance equations for currents at the left and right contacts.
  We find
    \aleq{
 \label{Pcl}
 &P_z=\frac{g_{ l} - g_{ u}}{2+   (g_{ u}+g_{ l})\coth \lambda}.
 }
Hence, the finite polarization exists even in the classical regime and is  therefore robust to dephasing \cite{conductance}.

\begin{figure}[h!] \includegraphics[width=0.9\columnwidth]{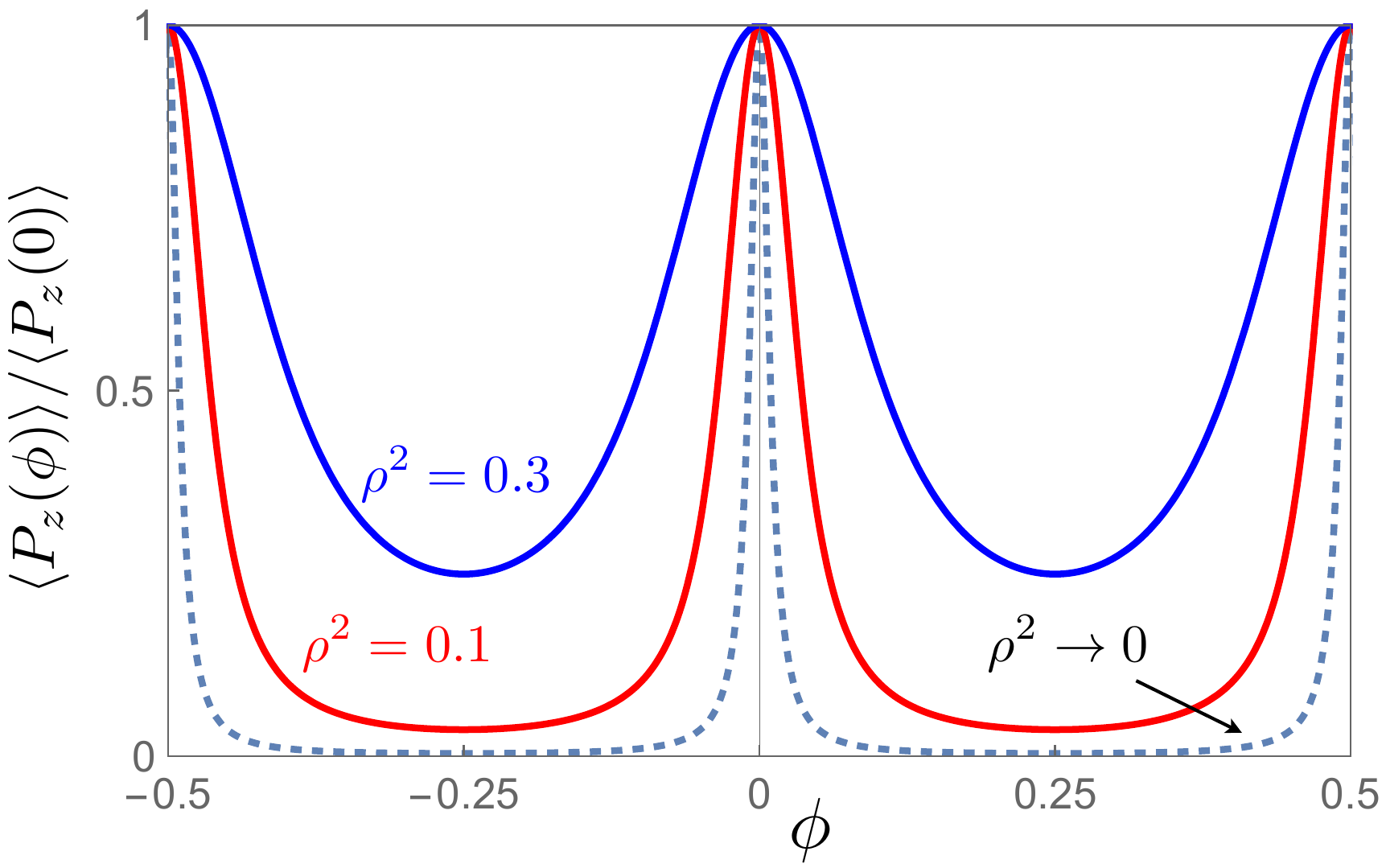}
\caption{\label{fig-disorder}
 (Color online) Broadening  of resonances  in polarization with increasing  strength of magnetic disorder,
$\rho^{2} = (N_{u} +N_{l})\rho_{0}^2/3$,
and  $\lambda=0.03$.}
 \end{figure}

The above perturbative analysis of a single impurity case shows that  all quantum effects  are encoded in the renormalization of backscattering probability: $\theta^2 \to {\tilde \theta}^2.$ Physically, it happens because such effects arise due to the interference of
multiple returns to magnetic impurity
along
the ballistic  trajectories propagating in   opposite directions  and  having  the same winding numbers.
Therefore, generalization for the case of many impurities is trivial: one should expand Eq.\ \eqref{Pcl} over impurities backscattering probabilities in   lowest order and  take into account the renormalization,  Eq.~\eqref{factor}.  For the case  of weak impurities of equal strength,  we find that $\mathcal T$ is given by Eq.~\eqref{T-pert} with the replacement  ${\tilde \theta}^2  \to {\tilde \theta}^2 (N_u+  N_l),$ and the  polarization reads
\aleq{
&  P_z= \frac{ {\tilde  \theta}^2(N_l-N_u)}{2} ~ \underset{\lambda \ll 1}{\longrightarrow}  ~ \frac{2 \lambda \theta^2 (N_l-N_u)}{1-\cos(4\pi \phi) +8\lambda^2 }    .
}
 One can  generalize this formula  in order to take into account non-perturbative effects  with respect to 
 impurity strength (still assuming $\theta<1$).    Corresponding calculations are presented in the Suppl. Material. The result is shown in  Fig.~\ref{fig-disorder}. As seen, non-perturbative effects lead to broadening of the resonances.       
 
 One of the most important conclusions of this section is universality of obtained results which was discussed  previously in context of conductance calculation \cite{my-cond}. The final equation for polarization is not sensitive to geometry of device and details of the structure. Also, the Berry phase drops out from the final result. 
 Physically, this happens due to our assumption $T \gg \Delta. $ In this case,  quantum contribution 
 to the conductance  depends on quantum return probability (ballistic Cooperon) which is the universal quantity.       

\subsection{Ensemble of qubits}
The transport through a HES-based interferometer was examined above (and
earlier in \cite{my-cond}) by a direct summation of the amplitudes of
quantum transitions.
Equivalently, the charge transfer through the interferometer can be viewed
as a tunneling through an ensemble of equivalent qubits.

The latter approach is applicable 
for important case of either $ \phi \ll 1 $  or $\phi-1/2 \ll 1$   and
weak impurities. Although it does not allow one to describe  transmission
coefficient and polarization for  $\phi \sim 1,$   it is more illustrative
physically  and much more suitable for the analysis of quantum computing
in the system under discussion. Below,  we  discuss this approach for the case if the interferometer 
with  a single magnetic impurity.

\begin{figure}[ht]
\includegraphics[width=0.9\columnwidth]{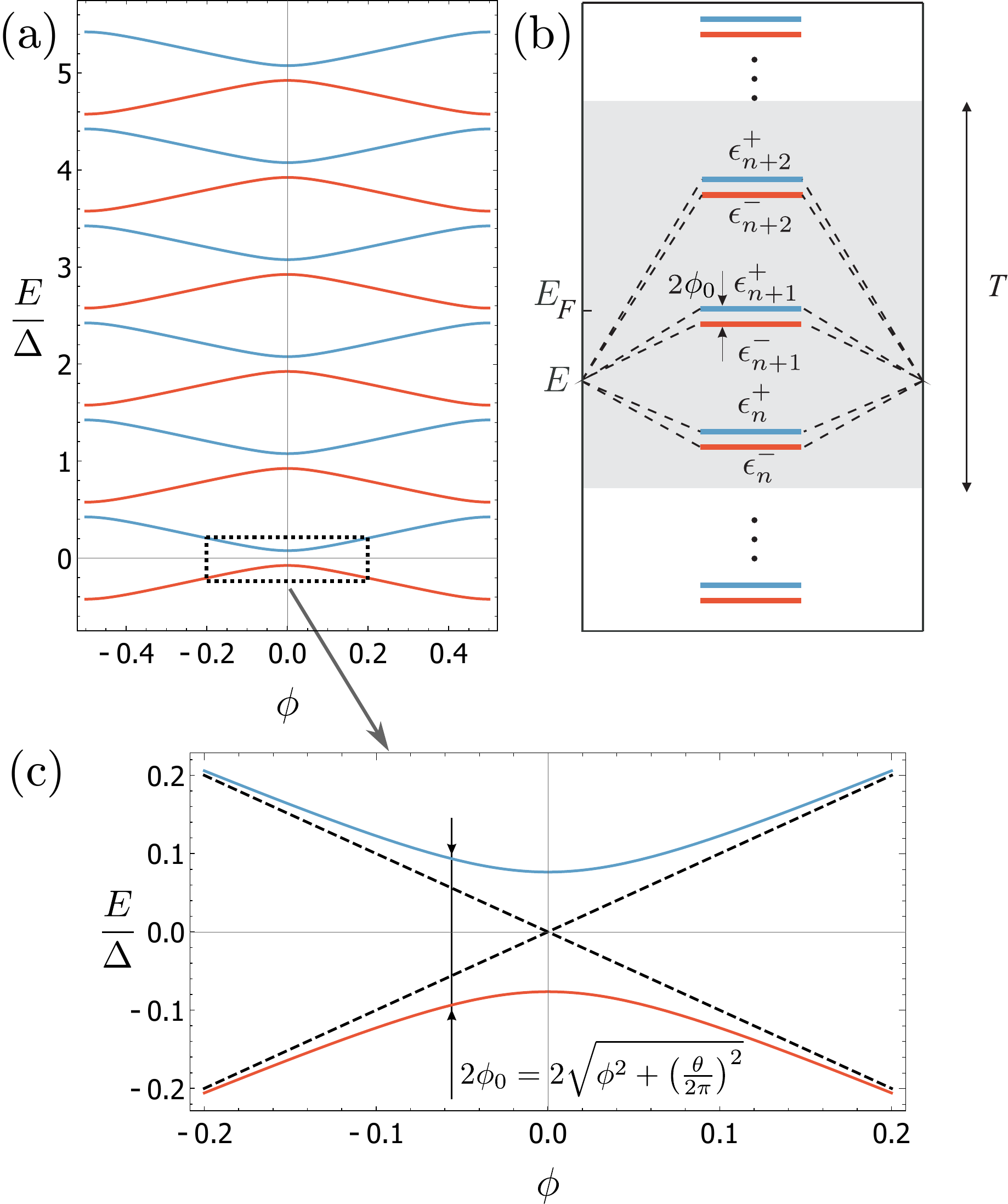}
\caption{
 (Color online)  (a) Energy levels  of right- and left- moving electrons (red and blue curves, respectively)    in the closed interferometer; (b)  Transmission of the  electrons through ensemble of $T/\Delta$ active qubits;  (c) Anticrossing at  $\phi=0.    $  }
\label{fig-levels}
 \end{figure}

The key idea is that  the tunneling amplitude through the interferometer
can be presented as a sum of the transition amplitudes through
intermediate states corresponding to quasistationary levels of an almost
closed HES (similar approach for non-helical single-channel interferometer
was discussed in Ref.~\cite{Shmakov2013}). As a starting point, we
consider an interferometer in the limit of an infinitely weak tunnel
coupling, i.e. a system of two closed HES. In the absence of magnetic
impurity, quantum levels are given by the  following formula,
$\epsilon^{\pm}_n(\phi)=\Delta(n \pm \phi),   $ and for integer and
half-integer values of the flux, the level system is degenerate:
$\epsilon^+_{n}(0)=\epsilon^-_{n}(0),$
$\epsilon^+_{n}(1/2)=\epsilon^-_{n+1}(1/2).$ Magnetic impurities lift
this degeneracy. In particular, for a single magnetic impurity described
by Eq.~\eqref{SM}, quantum levels are given by \be
\epsilon^{\pm}_n=\Delta(n \pm \phi_0), \label{energy pm} \ee where
$\phi_0$ obeys \be \cos (2 \pi \phi_0) = \cos \theta \cos  (2 \pi \phi) ,
\label{phi0} \ee hence, anticrossing at $\phi =0 $ and  $\phi= 1/2.$ The
energy levels are plotted in Figs.~\ref{fig-levels}~(a,b).  For weak
impurity, splitting at anticrossing points,
$(\epsilon^+-\epsilon^-)|_{\phi=0}=2\Delta \theta,$  is small.

The form of wave functions, provided in Suppl. Material, shows that 
   spinors
corresponding to different $n$ have the same direction of local spins  at
the impurity position:
$$ \mathbf S^\alpha (x_0)=- \mathbf S^{-\alpha}  (x_0) =   \frac{1}{2}
\langle \psi_n^ \alpha (x_0) |    \boldsymbol{ \hat  \sigma}  |\psi_n^ \alpha (x_0) \rangle, \quad \alpha = \pm . $$
With increasing $x$  starting from  $x=x_0 +0,$   $z$-component of local
spin does not change, $S_z^\alpha(x) =S_z^\alpha(x_0).$      By contrast,
the perpendicular  component of  local spin  rapidly rotates, rotating  by
angle $4 \pi (n \pm \phi_0) $  upon arrival  to the point $x=x_0 -0$ after
passage of the ring.

\begin{figure}[ht!] \includegraphics[width=0.9\columnwidth]{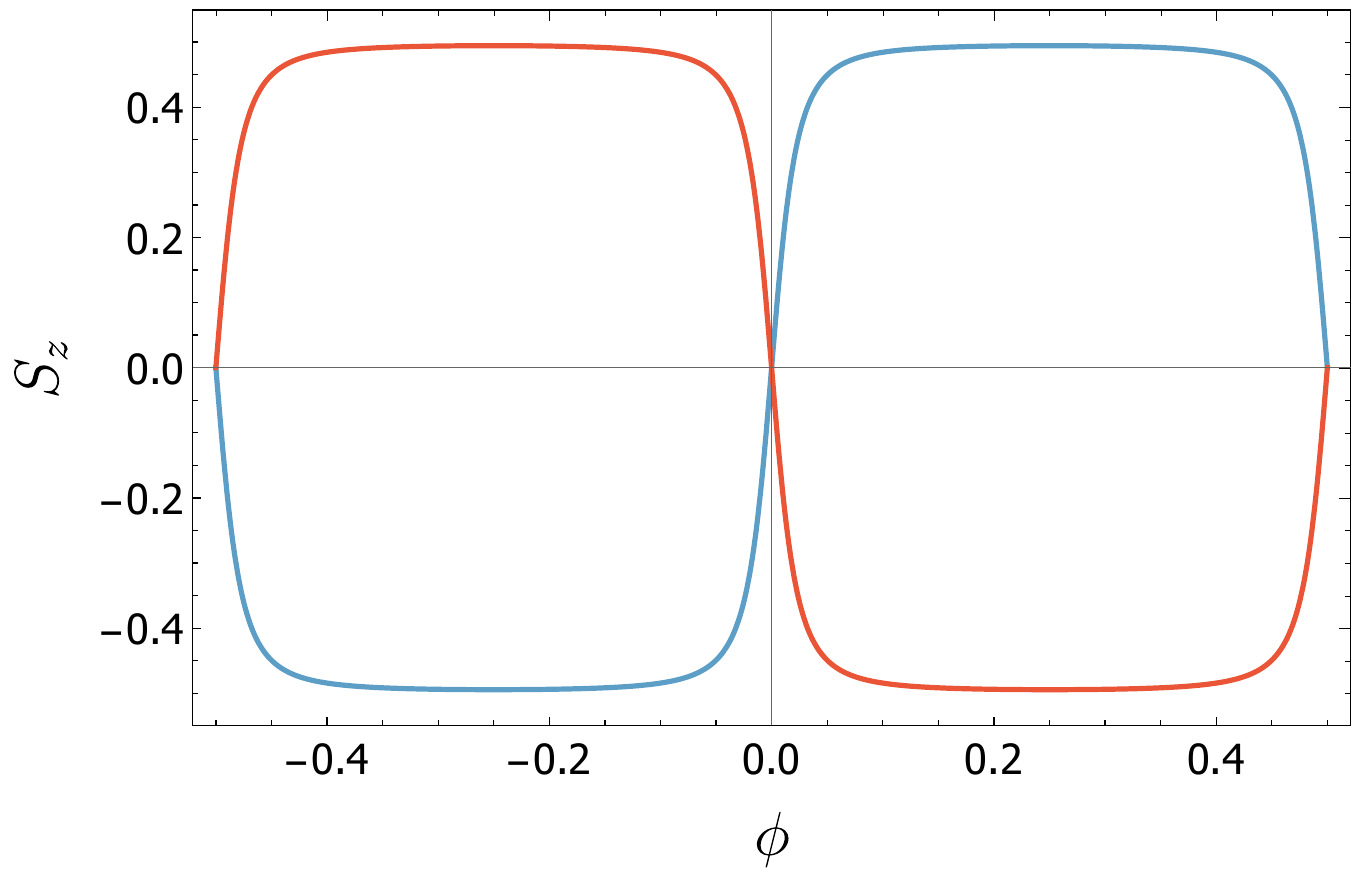}
\includegraphics[width=0.9\columnwidth]{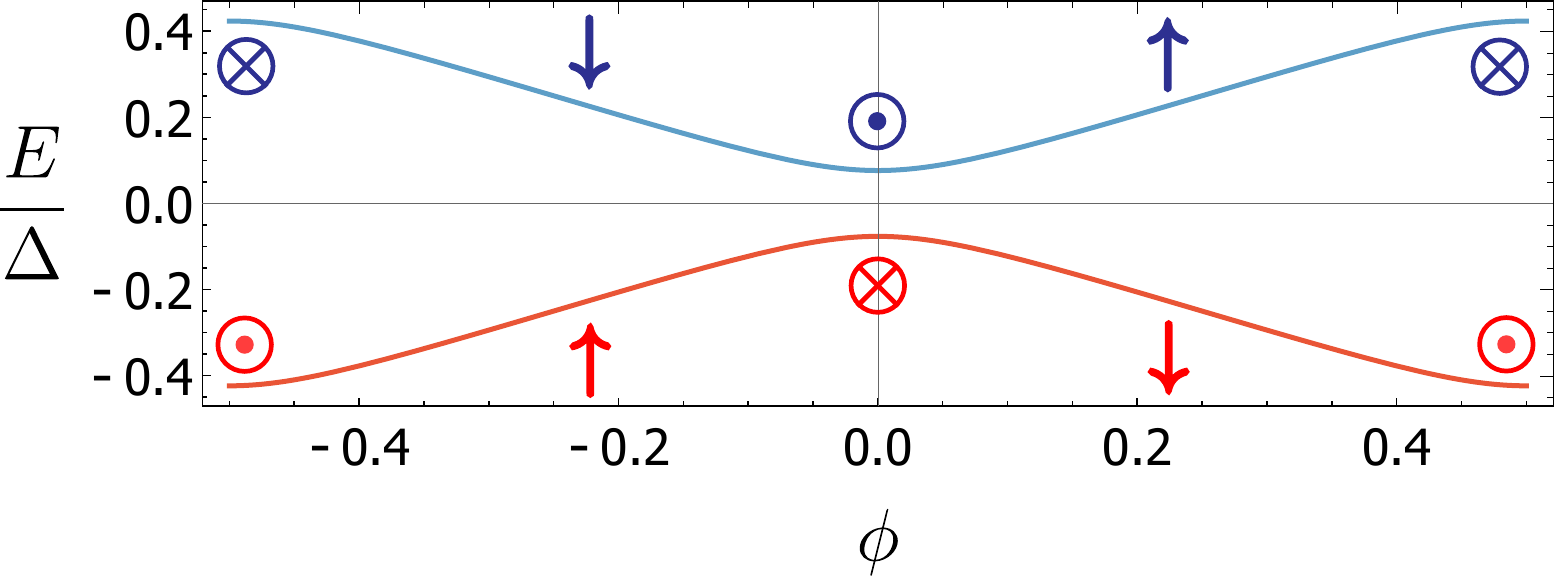}
\caption{
 (Color online) Variation of qubit states  with the magnetic flux. For $\phi=0$ both right- and left-moving electrons (red and blue curves, respectively)  have  spins perpendicular to the $z-$axis.  Electron spin switches between $S_z=1$ to $S_z=-1$ within narrow interval of $\phi.$}  
\label{spin_oscill}
 \end{figure}

Anticrossing at $\phi =0$ is illustrated in Fig.~\ref{fig-levels}~(c)
(picture at   $\phi =1/2 $ is fully analogous).  For weak impurity, in
vicinity of anticrossing point,  we have \be 2\pi \phi_0 \approx \sqrt{(2
\pi \phi)^2 +\theta^2,} \ee and, consequently, \be \delta \epsilon =
\epsilon^{+}_n - \epsilon^{-}_n  \approx  2 \Delta \sqrt{\phi^2+
(\theta/2\pi)^2}, \label{distance} \ee for $\theta\ll 1,~ \phi \ll 1.$ As
seen,  close to anticrossing points the distance between $(n,+)$ and
$(n,-)$ is small, so levels are almost degenerate,  and          can be
controlled either by perpendicular magnetic field, which effects both
$\phi$ and $\theta,$ or   by parallel  field, which also rotates moment of
the magnetic impurity thus changing $\theta.$
 Close to points  $\phi=0$ and $\phi=1/2$,  $z-$component of spin changes very sharply (see Fig.~\ref{spin_oscill})
 \be
 S_z^\pm  \approx  \mp  \frac{ \pi \phi}{\sqrt{ (2 \pi \phi)^2+ \theta^2}}, \quad \text{for}~|\phi| \ll 1,~\theta \ll 1,
 \ee
and similarly for $|\phi -1/2| \ll 1.$

For week tunneling coupling  ($\lambda \ll 1 $), the tunneling transport
of the electrons   through interferometer can be described in terms of
transmission amplitudes (see \cite{Shmakov2013}  and Suppl. Material)  \be  A_n^\alpha(\epsilon)   \propto
\frac{1}{\epsilon -\epsilon_n^\alpha - i \Gamma/2} + \dots, \ee where
$\Gamma\approx 2\Delta \lambda/\pi $  is the tunneling rate, and  $+ \dots$ stands for non-singular contribution.          
Both $\cal
T$ and $P$ 
can be expressed in terms of energy-averaged   bilinear  combinations of these amplitudes. There are   
``classical'' terms, $\propto \langle |A_n^\alpha(\epsilon)|^2 \rangle_\epsilon ,$ and interference terms, $\propto \langle A_n^\alpha(\epsilon)A_m^{\beta *}(\epsilon) \rangle_\epsilon $  with  $(n,\alpha) \neq (m,\beta),$  corresponding to transitions through different quantum levels   (see Fig.~\ref{fig-levels}b). For the case under discussion, $\lambda \ll 1,~T\gg\Delta, $  the interference contribution is dominated by   terms  with $n=m$ and $\beta=-\alpha,$
\be
\left \langle A_n^\alpha(\epsilon)A_n^{-\alpha *}(\epsilon)\right \rangle_\epsilon \approx\frac{2 \pi i}{\alpha \delta \epsilon + i \Gamma} \left(- \frac{\p f_F}{\p \epsilon }\right)_{\epsilon=n\Delta},
\ee
while  interference processes  with $n\neq m$ are described by similar equation which contain  term    $(n-m) \Delta \gg \delta \epsilon $  in the denominator and therefore is small.  
\subsection{Quantum computing  by qubit  ensemble}
It is known  that  conventional interferometers with spin-orbit (SO) interaction (or an array  of such interferometers) can be used as one-qubit quantum gates of various types (X-gate, Z-gate, phase gate, and Hadamard gate) \cite{Foldi2005}, which manipulate spin states of the electrons with given energy---the so-called flying qubits.  The flying qubits can be used for quantum calculations at very low temperatures $< 100$ mK \cite{Baurle2018}.     Analyzing  analytical expression for  
energy- and spin-dependent  transmission amplitudes $t_{\alpha \beta}(\epsilon)$ [see Eq.~\eqref{tab} of the Suppl. Material]    
one can---in a full analogy with Ref.~\cite{Foldi2005}---introduce quantum gates of different types.  However, 
here we would like to focus on a different issue, namely, possibility of high-temperature qubit manipulation.     Since, we consider almost closed  tunneling interferometer,  we will use language of the quantum levels  introduced in the previous section.  

The almost degenerate  pairs of levels represent an  ensemble of qubits
with equal interlevel distance.  The number of active qubits, which are
able to  participate in the spin and charge transport is given by \be
{\cal N}_{\rm active} \simeq \frac{T}{\Delta}. \ee Transmission of charge
and spin through the interferometer can be considered  in terms of
coherent hopping through these qubits  (analogously to the case of
conventional interferometer  \cite{Shmakov2013}) as illustrated in Fig.~\ref{fig-levels}b.

{Technically, in order to describe  transition though qubit levels one should introduce projection operators $\hat P_1$ and $\hat P_2$ (see Suppl. Material), which can be presented as  
$$P_{1,2} = \frac12 (1 \pm \hat H)\, .$$ } 
Here we introduced Hadamard operator  
\be \hat H =
\begin{pmatrix} a & b  e^{ - i \xi}
\\ b e^{  i \xi}  & -a \end{pmatrix},
\label{Hadamard11}
\ee
where coefficients
$b= { e^{-2 \pi i \phi} \tan\theta}/{ \sin (2\pi \phi_0) }$ 
and $a= i\left[
e^{-2 \pi i \phi}/\cos \theta -\cos(2\pi \phi_0)  \right] / 
\sin(2 \pi \phi_0) $,
 obey $a^2+b^2=1$ and depend on the strength of the impurity and the magnetic flux only, while
 the dependence on the   energy is encoded in the exponents
 $e^{\pm i \xi}$ entering off-diagonal terms of $\hat H.$
The operator 
$\hat H$ 
 has   standard properties
 \be
 \hat H^2=1, \quad {\rm Tr} ~\hat H =0, \quad {\rm det} ~\hat H=-1. \ee
Importantly,
 $\hat H$ can be tuned by the external magnetic field.

Off-diagonal elements of $\hat H$ rapidly oscillate with energy and, strictly speaking,
  one could  introduce  a set
    of   Hadamard operators corresponding to different quantum levels in the interferometer:
 $\hat H_{n\alpha} =\hat H_{\epsilon=\epsilon_{n\alpha}  }.$  However, the results of direct calculations  for conductance  and spin polarization  show that the dependence on $n$ drops out.   Hence,   we  have an ensemble of qubits,  which give  coherent contributions to the charge and spin transport.        

 For $\theta \ll 1, \lambda \ll 1,  $ and  $\phi \ll 1,$ the transmission coefficient and polarization are expressed in terms of
$\hat H$ as follows (see Suppl. Material)
\be
\mathcal T \approx \frac{\pi \Gamma}{8 \Delta} {\rm Tr}( \hat A), \qquad    { P}_z\approx \frac{\pi \Gamma}{8 \Delta \mathcal T} {\rm Tr}( \hat \sigma_z \hat A), 
\label{T-P-had}
\ee
where   $\Gamma= 4 \lambda v_F/L = 2 \lambda \Delta/\pi$ is the tunneling rate and 
\be
\hat A=  \frac{\hat S (\Gamma+ i \delta \epsilon\, \hat H )^\dagger
\hat S (\Gamma+ i \delta \epsilon\, \hat H )}{\Gamma^2+ \delta \epsilon^2}+\frac{2\pi \Gamma}{\Delta} \hat \sigma_z,
\ee
where information about tunneling coupling is encoded in   $\Gamma$ and in the matrix
\be
\hat S= \begin{pmatrix} e^{-\lambda} & 0  
\\ 0  & e^{\lambda} \end{pmatrix}. 
\ee
Hence, measurement  of $\mathcal T$ and $ P_z$ allows one to read out information about ensemble of qubits.  Importantly, the results of calculation do not depend on energy (entering through factor $\xi$).  In other words, all qubits give equal contributions to conductance and polarization. 

Using Eq.~\eqref{Hadamard11},  for small  $\theta,  \lambda ,  $ and  $\phi ,$ we get
\be
 \mathcal T\! \approx\! \lambda\! -\! \frac{ \lambda^3 \theta^2}{4 \lambda^2\!+\! \theta^2\! +\! 4\pi^2 \phi^2},
~ P_z\!\approx\!  - \frac{ \lambda \theta^2}{4 \lambda^2\!+\! \theta^2\! +\! 4\pi^2 \phi^2}
\label{T-P-final}
\ee
The same equations are valid for $\phi$ close to $1/2$ with the replacement $\phi \to\phi-1/2. $ 
One can check by direct calculation that dependence on $\xi,$ and, consequently, on energy drops out after taking trace in Eqs.~\eqref{T-P-had}.    We see that the approach based on qubit representation not only  reproduces results obtained by direct summation of the amplitudes within $\theta^2$ precision but also allows one to perform non-perturbative summation over relevant scattering processes and to get $\theta^2$ in the denominator of Eqs.~\eqref{T-P-final}.   Sharp dependence  of  $\mathcal T$ 
and $ P_z$ on $\phi$ reflects tunability of the ensemble of qubits by external magnetic field.

\begin{figure}
\includegraphics[width=0.5\columnwidth]{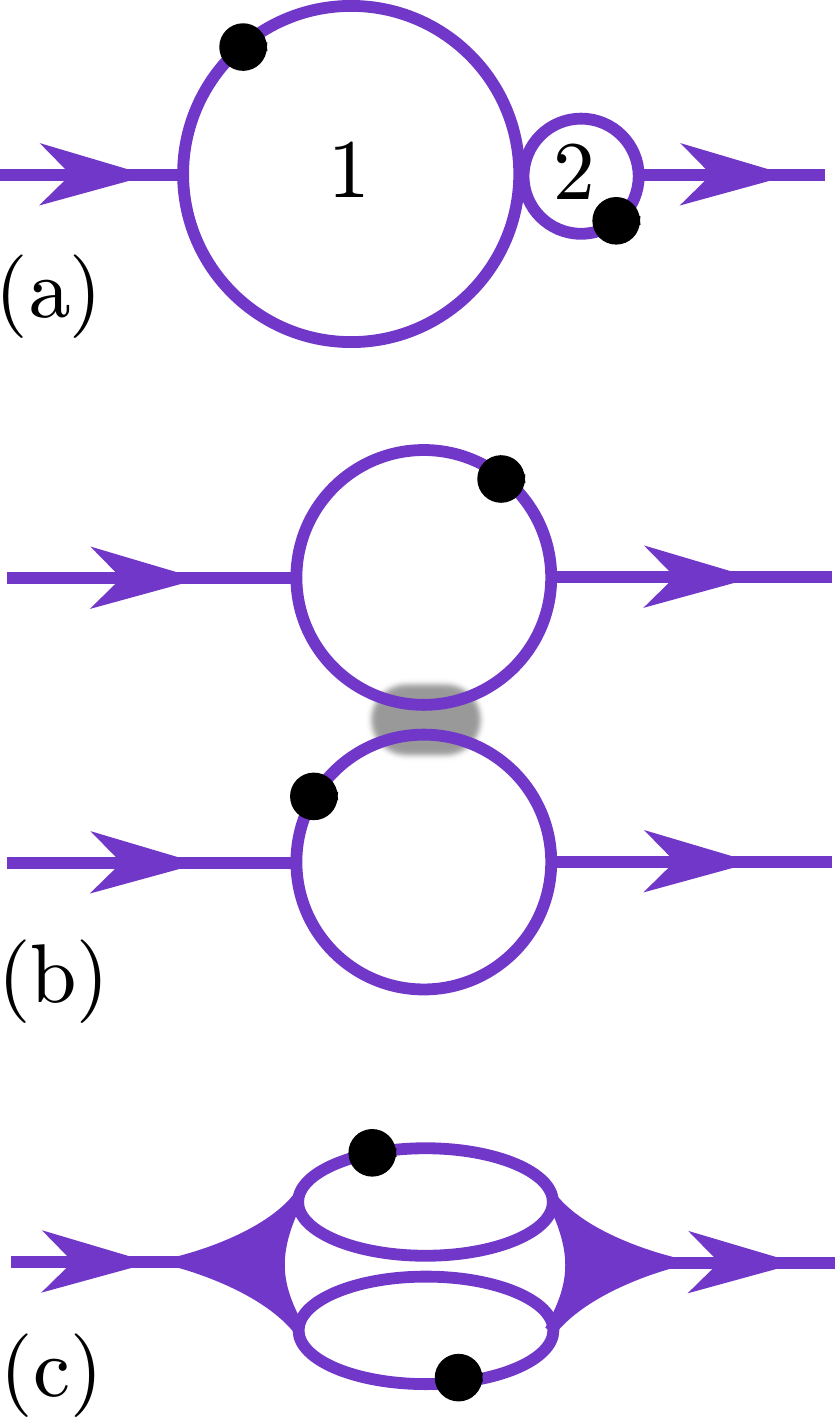}
\caption{
 (Color online) Several setups containing   two  AB interferometrs: (a)     Setup for one-qubit operations based  on interferometers of different sizes connected in series. Interferometer of larger size is  more sensitive to magnetic field and is used  to manipulate the qubit spin state in the interferometer of smaller size (see also Fig.~\ref{energy2rings});     (b) Setup for two-qubit operation based on 
    interferometers connected in parallel and coupled by   electron-electron interaction (the interaction region is  
    marked by grey color); (c)  Setup for creation of pure state of arbitrary polarization with two  interferometers containing strong impurities that block transmission through corresponding   shoulders of each interferometer and with  
    joint contacts to metallic leads allowing  for 
    coherent tunneling  to both interferometers.               }
\label{doublerings}
 \end{figure}
 
 \begin{figure}
 \includegraphics[width=0.7\columnwidth]{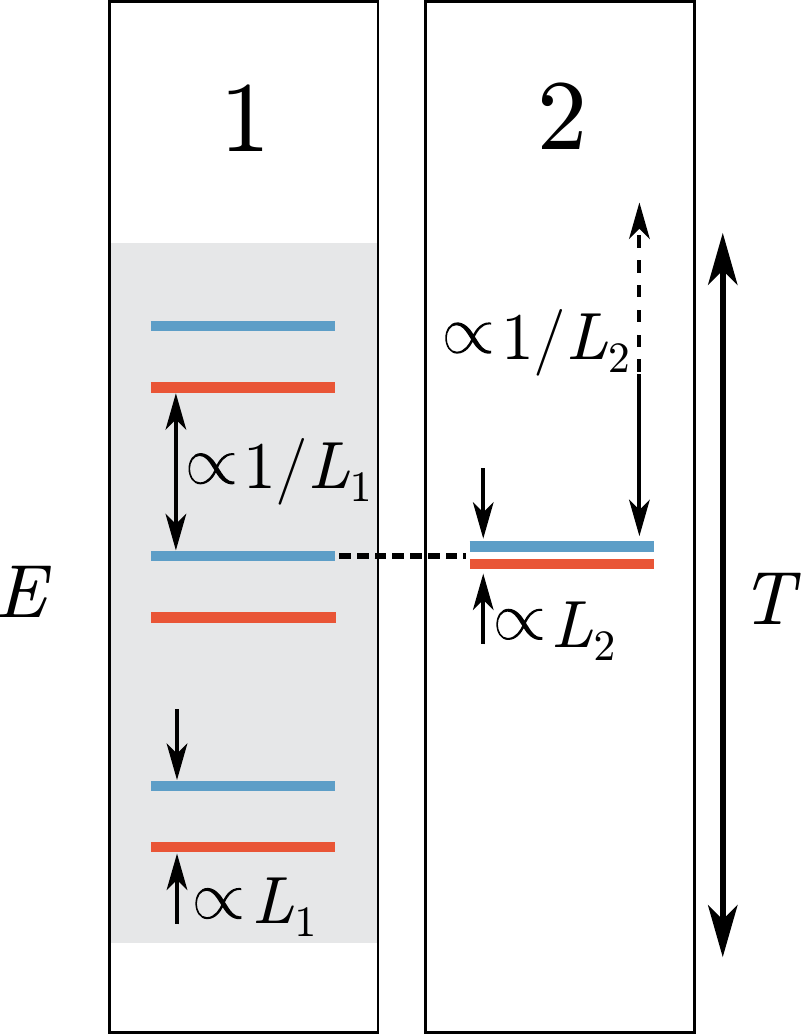}
\caption{
 (Color online) Energy levels in setup with two interferometers of different sizes connected in series (see Fig.~\ref{doublerings}a). Energy levels in the interferometer of larger size are much more sensitive to magnetic field and can be tuned to manipulate  a polarization of a single active qubit in the interferometer of smaller size.  }
\label{energy2rings}
 \end{figure}

It is interesting to discuss possible generalizations of the high-temperature computing schemes   to more complex systems involving several interferometers based on  HES or   HES arrays. 
(Experimental study of HES arrays has recently begun
\cite{Maier2017}.)  
The simplest examples of setups  with two interferometers are shown in Fig.~\ref{doublerings}.  
Figure ~\ref{doublerings}a schematically depicts two   interferometers tunnel-connected in series  to leads and to each other, with different edge lengths $L_1$ and $L_2$ ($L_1 \gg L_2$).  In the absence of magnetic field, level spacings  in these interferometers   are very different: $\Delta_1 \propto 1/L_1 \ll \Delta_2 \propto 1/L_2 .$   Then,  for
$ \Delta_1 \ll T \ll \Delta_2,$  there are $T/\Delta_1$ active qubits in the first interferometer and single active qubit in the second one   (see Fig.~\ref{energy2rings}). On the other hand,   for weak impurities,  spacing between qubit's levels is much larger in the first interferometer, $2\Delta_{1} \phi_{1} \propto L_1 \gg  2 \Delta_2 \phi_2 \propto L_2 $ (here, we assume that homogeneous magnetic field is applied to both systems, so $\phi_{1,2} \propto L_{1,2}^2$). Hence, the first interferometer is much more sensitive to magnetic field.  In particular,  one can tune  an energy level in the system 1 to be in the resonance with the levels of active qubit in the system 2. Then, one can change the pure quantum state of the qubit 2 by very small variation of the external field.      

Similar to the low-temperature case  \cite{Bautze2014,Baurle2018,Bordone2019,Bellentani2020,Chen2014} one can suggest 
two qubit manipulation  schemes taking into account the electron-electron interaction. To this end, one can use   interferometers connected in parallel and coupled by    interaction (see Fig.~ \ref{doublerings}b). The most essential feature of the high-temperature  case distinguishing it from the  low-temperature one is that now effective manipulation is possible for the whole ensemble of qubits.  In particular, simplest capacitive interaction between two interferometers would lead to the respective interaction-induced phase  shift between  states in the  upper and down systems.       

Finally, one can construct a   setup for creation of outgoing  polarized  state with arbitrary polarization  direction by using 
two  interferometers containing strong impurities that block transmission through corresponding   shoulders of each interferometer and with   
    joint contacts to metallic leads allowing  for 
    coherent tunneling  to both interferometers (see Fig.~ \ref{doublerings}c).   Assuming that unpolarized   
    electrons enter the system from the left contact, we find that at the right contact there is  interference of two 
    pure coherent states with various (in general, arbitrary) polarizations. As a result, outgoing electrons will be  polarized with the direction different from outgoing polarization of each interferometer.

\section{Conclusions}

We have studied coherent spin transport through HES of 2D topological
insulators. We have shown that unpolarized incoming electron beam entering
the HES through one of the metallic leads acquires a finite polarization
after transmission through the setup containing magnetic impurities. The
finite polarization appears even in the fully classical regime and is
therefore robust to dephasing.   
There also exists quantum contribution  which survives at relatively high temperature and is tunable
by magnetic flux piercing the area encompassed by HES. Specifically, the
quantum contribution shows sharp identical AB resonances as a function of
magnetic flux with maxima  (in the absolute value) at integer and
half-integer values of the flux. For the setup with   a single strong
magnetic impurity blocking the transmission in one shoulder of AB
interferometer, and for large tunneling coupling, the spin polarization of transmitted electrons can achieve
100\%, which implies  that outgoing electrons are in the pure quantum spin state. 
Also this means that polarization can be transferred  over distances on the order of the
system size. The polarization reverses sign when
impurity is moved from one shoulder of interferometer to another. 

  We discuss possible application of obtained results for quantum computing. We demonstrate that  tunneling interferometer based on HES can be described in terms of ensemble of flux-tunable   qubits giving equal contributions to conductance and spin polarization.  Specifically, 
  in presence of magnetic impurities and magnetic field the initially doubly degenerate HES spectrum is split so that the appearing pairs of quantum states act as qubits with the spin orientation easily tuned by magnetic flux.  The number of active qubits participating in the charge and spin transport is given by the ratio of the temperature and the level spacing. The interferometer can  effectively operate at high temperature and can be used for quantum calculations.  In particular,  the ensemble of  qubits  can be described by a single flux-tunable  Hadamard operator.      These findings  are not sensitive to details of the system such as geometry of the HES and  allows one to speak about single-qubit operations such as X or Z gate. Since we  also predict the polarized state after passing the AB interferometer by the unpolarized beam, we can   prepare the qubits in the desired states. 
  
  If one uses the outgoing polarized state  as the input for the next AB interferometer, then one can further manipulate the states of the qubits. Arranging the setups involving several interferometers of  certain geometries we can produce non-trivial two-qubit operations needed for quantum computations.  
  The obtained results open wide avenue for applications in the area of quantum computing.

\section{Acknowledgements}
 The work  was supported by  the Russian Science Foundation
(Grant No. 20-12-00147)
 and   by   Foundation for the Advancement of Theoretical Physics and Mathematics ``BASIS''.
 Work in  Poland was supported by  the Foundation for Polish Science through the grant MAB/2018/9  for CENTERA.

\newpage
\begin{widetext}

\hspace{5cm} {\huge{ Supplemental material}}
\setcounter{section}{0}

\begin{quote}
In this Supplemental Material, we provide a short discussion of the Rashba coupling effects, derive an analytical expression for the transfer matrix of the interferometer,  and analyze the spin polarization for the case of a large number of weak, randomly distributed magnetic impurities.
\end{quote}



\section{Rashba coupling}
The Rashba coupling is described by the following term in the Hamiltonian
\be
{\cal H}_{\rm Rashba}= \sum \limits_{\alpha,\beta=\uparrow, \downarrow}
\psi_\alpha^\dagger {\sigma}_{\alpha \beta}^y \{  a(x),  i \partial_x\} \psi_\beta,
\label{HR}
\ee
where $ a(x)\, \mathbf n $ is the 
local  Rashba field,    $\mathbf n$ is  the  unit vector perpendicular   to the plane of the topological insulator, and 
$\{ \cdots \}$ stands for anticommutator 
\cite{Kimme2016}. 
Assuming that the edge
is smooth at the scale $p_F$ and $a(x) \ll v_F$, we can use the
semiclassical arguments and integrate Schr\"odinger equation, corresponding
to  the  Hamiltonian of HES,   exactly. Such analysis was
performed in Ref.~\cite{Shmakov2012} for conventional (non-helical) materials  and showed the appearance of Berry phase upon the
whole revolution around the edge.   It was shown however in Ref.~\cite{my-cond}
that for our purposes and thanks to nature of helical edge states, the
Berry phase is irrelevant. 

\begin{figure}[h!] \includegraphics[width=0.5\columnwidth]{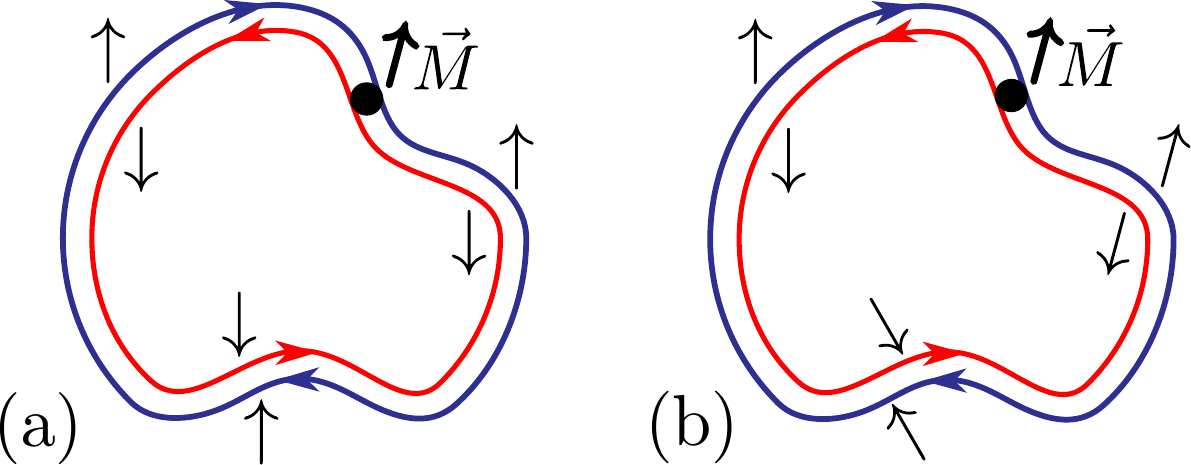}
\caption{  Helical edge states without (a) and with (b)  Rashba coupling.
 (Color online)  }\label{curveringimpuritiesRashba}
 \end{figure}

Let us calculate the phase acquired by  the electron wave  after  a full revolution around the setup shown in Fig.~\ref{fig:densityring}.   In order to make the physical picture more transparent,   we  consider first a  general case with two chiralities (direction of propagation) and two spins not necessarily aligned  with momentum (this case corresponds to a conventional single-channel spinful wire).
The  acquired phase
includes three terms:  a dynamical contribution $kL$  (here $k$  is  the electron wave vector),
magnetic phase $\pm 2\pi \phi$ and the Berry's phase $\pm \delta$~\cite{berry}, given by one half of the solid angle, subtended by spin direction during circumference of the interferometer.     The dynamical contribution depends on $L$ only and does not change its sign when changing the chirality  and spin projection.   By contrast,  the sign of magnetic phase is insensitive to spin but changes sign with changing the chirality. The Berry's phase changes sign both with changing the chirality and with changing the  spin (see Tab.~\ref{tab:phases}).
 For helical edge only two out of four electron states are present, which are marked by boldface in the Table~\ref{tab:phases}.

\begin{table}[ht]
\begin{tabular}{cccc}
& & \multicolumn{2}{|c|}{chirality}   \\
\cline{3-4}
& & \multicolumn{1}{|c|}{+} &\multicolumn{1}{c|}{--} \\
\hline
{spin}  
& \multicolumn{1}{|c|}{$\uparrow$} & \multicolumn{1}{|c|}{ \boldmath$k L + \phi + \delta$}  & \multicolumn{1}{c|}{$k L - \phi - \delta$ }  \\ \cline{2-4}
 & \multicolumn{1}{|c|}{$\downarrow$} & \multicolumn{1}{|c|}{$k L + \phi - \delta$}  & \multicolumn{1}{c|}{ \boldmath$k L - \phi + \delta$} \\ \hline
\end{tabular}
\caption{\label{tab:phases} Phases of electron wave function after a full revolution in the arbitrary setup shown in Fig.\  \ref{fig:densityring}. }
\end{table}

 Analyzing  the corresponding phases we arrive at a conclusion, which is of
 key importance for our  analysis.
 Information about  the geometrical structure of the edge states, in
 particular, about  curvature of the   edge and/or non-planar geometry, is
 encoded in the Berry's phase, but as we see, it
    is  simply   added to the dynamical phase, which implies  that
    amplitude  of  any process depends on $kL+\delta.$ This, in turn,
    means that tunneling conductance for a given energy (i.e. before
    thermal averaging) depends on the Berry's phase and is, therefore,
    sensitive to geometry of  the setup. However,  for $T\gg \Delta,$  the
    thermal averaging  implies integration over $k $  within a   wide
    interval, $ \delta k \sim T/\hbar v_F \gg 1/L,$  around the Fermi wave
    vector  $k_F.$ After changing integration variable,   $k+\delta/L \to
    k^\prime$,  the Berry's phase drops out with  the exponential
    precision.    This should be contrasted to the case of conventional
    interferometers with weak SO coupling, where the Berry's phase
    contributes to the Aharonov-Casher phase and  strongly effects both
    $\mathcal T(\epsilon)$  and     energy-averaged transmission
    coefficient, $\mathcal T$~\cite{Shmakov2012}.   Physically, this happens because for weak
    SO coupling, the   electron wave with a given spin polarization can
    propagate  both clockwise and counterclockwise and  the  phase shift
    between such waves   with equal winding numbers, $n_1=n_2=n$, is given
    by $2(\phi+\delta) n$.

The conclusion formulated above requires a minor comment. As seen from the Fig.~\ref{curveringimpuritiesRashba},  
spin rotates while an electron passes the interferometer. The parameter $\theta$ which determines the scattering strength depends on the direction of magnetic moment of the impurity with respect to the local spin quantization axis. The direction of outgoing polarization is also parallel to the local quantization axis at the position of outgoing contact.

\section{Transitions through energy levels of closed ring }
Here we derive analytical expressions describing anticrossing of quantum levels of  right- and left-moving electrons on the example of a single impurity placed in the upper shoulder.
We consider interferometer with the lengths of  the upper and lower  shoulders given by $s$ and $L-s,$ respectively.  The magnetic impurity is placed at position $x_0$ such that     $0<x_0<s.$
Using expression for scattering matrix \eqref{SM}, one can easily find  transfer matrix of  impurity 
\be \hat W= \frac {e^{-i \alpha }}{\cos \theta}\left(   \begin{array}{cc}
  1 &  i \sin \theta e^{-i \xi} \\
      -i \sin \theta e^{i \xi} & 1 
  \end{array}
   \right),
     \label{eq:W}
     \ee
 where $\xi= \varphi+ 2 k x_0.$ In this supplementary we may add the constant value of forward scattering phase $\alpha$   to the flux $2\pi \phi$ and set $\alpha=0$ below.
 The solution of the scattering problem for the electron with momentum $k$ on the whole system yields 
 \be
 \begin{pmatrix}
 a^\uparrow  \\
                   a^\downarrow 
                                              \end{pmatrix}
                                          =  \hat t  
                            \begin{pmatrix}                 
                                                b^\uparrow  \\
                                                 b^ \downarrow 
                                            \end{pmatrix}
 \ee
where  $(b^\uparrow, b^\downarrow)$  and   $(a^\uparrow, a^\downarrow)$ are the amplitudes of incoming (from the left contact) and outgoing (to the right contact) waves and 
\be \label{tab}
\hat t=  (1-e^{- 2\lambda})  e^{2 \pi i \phi s/L} \begin{pmatrix} e^{i k s} & 0 \\ 0&e^{-i k s} \end{pmatrix} \begin{pmatrix} 1& 0 \\ 0&e^\lambda \end{pmatrix}\hat g \begin{pmatrix} 1 & 0 \\ 0&e^{\lambda} \end{pmatrix},
\ee
 where 
  \be
 \hat g= \frac{1}{1- \hat W \hat \Lambda } \hat W  \begin{pmatrix} 1 & 0 \\ 0&-1 \end{pmatrix},
 \ee
 \be
 \hat \Lambda=\begin{pmatrix} e^{i( QL + 2\pi \phi)} & 0 \\ 0&e^{i(- QL + 2\pi \phi)} \end{pmatrix},
 \ee
 and $Q$ is found from the condition $t^2 e^{i kL}=e^{i Q L},$ yielding
 \be
 Q=k+ i \frac{2 \lambda}{L}.
 \ee
The transmission coefficient and the spin polarization are expressed in terms of matrix $\hat t$ as follows
\begin{align}
&\mathcal T = \frac12 \left \langle {\rm Tr} \left(\hat t  \hat t^\dagger \right) \right \rangle_\epsilon,      
\\
&\mathcal P=\frac{1}{2 \mathcal T}\left \langle {\rm Tr} \left( \hat t \sigma_z  \hat t^\dagger \right)\right \rangle_\epsilon,
\end{align}
where $\langle \dots \rangle_\epsilon$ stands for  thermal averaging. Here  we neglect the Rashba coupling  and assume that the incoming electrons  are unpolarized. 
The matrix $\hat g$ can be presented as follows
\be
\hat g  = \cos \theta \left[  \frac{\hat P_1}{1-e^{i( QL + 2\pi \phi_0)}} +\frac{\hat P_2}{1-e^{i( QL - 2\pi \phi_0)}} + \begin{pmatrix} 0 & 0 \\ 0 & -1   \end{pmatrix}\right] ,
\label{expansionP12}
\ee
where $\phi_0$ is found from
\be
\cos(2\pi \phi_0)=\cos\theta \cos(2\pi \phi)
\ee
and 
\be
\hat P_1=\frac{1}{2 i \sin (2 \pi \phi_0) \cos \theta}\begin{pmatrix}-e^{-2i \pi \phi} +e^{2i \pi \phi_0} \cos \theta  & i e^{-i(\xi+2 \pi \phi)} \sin \theta \\i e^{i(\xi-2 \pi \phi)} \sin \theta  &    e^{-2 i \pi \phi} -e^{-2 i \pi \phi_0} \cos \theta  \end{pmatrix},
\ee
\be
\hat P_2=-\frac{1}{2 i \sin (2 \pi \phi_0) \cos \theta}\begin{pmatrix}-e^{-2i \pi \phi} +e^{-2i \pi \phi_0} \cos \theta  & i e^{-i(\xi+2 \pi \phi)} \sin \theta \\i e^{i(\xi-2 \pi \phi)} \sin \theta  &    e^{-2 i \pi \phi} -e^{2 i \pi \phi_0} \cos \theta  \end{pmatrix}.
\ee
These are projection operators obeying:   $\hat P_1^2=\hat P_1,~ \hat P_2^2= \hat P_2,~ \hat P_1  \hat P_2=0,~\hat P_1+ \hat P_2=1.$  Due to these properties, we can introduce  Hadamard operator
\be
\hat H = \hat P_1-\hat P_2,  
\label{H0}
\ee
which obeys the standard  property $\hat H ^2 =1.$
However, in contrast to conventional case,  we have $H \neq H ^\dagger.$ 
We can now write  
  \be 
\hat P_1= \frac{ 1+  \hat H}{2}, \quad \hat P_2=  \frac{ 1-  \hat H}{2}. 
\label{P12}
\ee
Thus defined Hadamard operator describes the isolated system and does not contain any information about tunneling coupling.
We use now the following identities  valid for arbitrary complex number $z$ with $\mbox{Im\,}z>0$ and 
arbitrary $\chi \in [0,1)$:
\be
\frac{e^{i \chi z}}{1- e^{i z} }= \left\{ \begin{array}{c}  
i \sum \limits_{n=-\infty}^{\infty}\frac{ e^{2 \pi i \chi  n} }{z- 2 \pi n}, \quad \text{for}~ 0<\chi<1 ,\\    
 i \sum \limits_{n=-\infty}^{\infty}\frac{ 1 }{z- 2 \pi n} + \frac12, \quad \text{for}~ \chi=0  \end{array}\right.
\label{chi}
\ee
%
Using Eqs.~\eqref{P12} and \eqref{chi}, we get
%

\begin{align}
&\hat g= \frac{i \cos \theta}{ 2} \left[ -  i \hat \sigma_z  +  \frac{\Delta}{ 2 \pi } \sum \limits_
{  n, \alpha}
\frac{1 + \alpha \hat H}{ \epsilon -  \epsilon_n^\alpha + i \gamma /2}   \right ],
\end{align}
with $\gamma= 4 \lambda v_F/L = 2 \lambda \Delta/\pi.$

 For completeness, we provide here the explicit form  of wave functions  for the energy levels 
 \eqref{energy pm}
\begin{align}
&\psi_n^\pm(x) = \frac{1}{\sqrt{|A^\pm|^2 +|B^\pm|^2}}\left[\! \begin{array}{c}
  e^{i  k_n^\pm (x-x_0) } A^\pm \\
  e^{-i  k_n^\pm (x-x_0) } B^\pm   \end{array}\! \right]
  \end{align}
Here, $\pm$ labels energy levels  [see Eq.~\eqref{energy pm}], $ k_n^\pm= \epsilon^{\pm}_n/v_F, $
 $A^\pm=\sin \theta e^{-i (\varphi  \pm 2\pi \phi_0) },$ $B^\pm= \cos\theta \sin(2 \pi \phi) \pm \sin(2\pi \phi_0),$  and $\varphi$
is the angle describing position of  the  magnetic moment of the  impurity
with a fixed projection on the local electron spin.
 Orthogonality condition reads
 $$\int \limits_0^L   dx \langle \psi_n^ \alpha (x)  |\psi_m^ \beta (x) \rangle=  \delta_{nm} \delta_{\alpha\beta} , \quad \alpha,\beta =\pm $$
We emphasize  that coefficients $A^\pm$ and $B^\pm$ that determines
direction of local spin at $x=x_0$   do not depend on $n.$

\section{Averaging over  positions of impurities}
\label{sec:averaging}

Next, we find non-perturbative expressions for  spin polarization 
    assuming that  $\rho_{0}^2 N_u \ll 1,~ \rho_{0}^2 N_l \ll 1 .$ In this case, the mean free path is much larger than $L,$ so that the regime is ballistic  and  one can neglect localization effects.
    We consider shoulders of equal length and replace interaction with $N_u$ ($N_l$)  impurities in the upper (lower) shoulder by    transfer matrix $\hat W_u$ ($\hat W_l$)
    describing scattering on the shoulder as a whole. In the ballistic regime,  parameters of this matrix
read
  \begin{equation}
  \theta_u e^{i \varphi_u}= \rho_{0} \sum \limits_{n=1}^{N_u} \sin \eta_n e^{i\varphi_n -2ik  x_n},\quad
  \alpha_u = \rho_{0}   \sum \limits_{n=1}^{N_u} \cos \eta_n
  \label{theta-alpha}
  \end{equation}
  (and $u \to l$ for lower shoulder).   We average the final polarization  over directions of vectors $\mathbf M_n,$  which means averaging over $\varphi_n$ and  $\eta_n.$  The  parameters of transfer matrix depend on
   positions of impurities, $x_n$. These positions, however, can be incorporated into  $\varphi_n$ and
  drop out  after averaging.

Let us consider interferometer with two effective impurities in the upper and lower shoulder, characterized by transfer matrices $\hat W_u = \hat W_u (\theta_u, \varphi_u, \alpha_u) $   and   $\hat W_l = \hat W_l (\theta_l, \varphi_l, \alpha_l), $ respectively. Assuming that $\theta_{u,l} \ll 1$ and in the  vicinity of the resonances,   $\delta \phi = \phi - n \ll 1 $ or $\delta \phi = \phi - (n+1/2) \ll 1$, we find the disorder-averaged spin polarization,    $\overline{P_z}  = \int P_z f_u f_l ~ du\, dl,$ where
    \be   P_z\!=\!\frac{\lambda (\theta_l ^2-\theta_u^2)}{ 4\lambda^2\! +\!    {(2\pi \,  \delta\phi\! +\! \alpha_u\!+\! \alpha_l)^2\! }+|\theta_u e^{i\varphi_u}\! -\!\theta_le^{i\varphi_l}|^2 }\,.    \label{P-two-imp}  \ee
and $f_{u}$, $f_{l}$ are distribution functions for parameters of matrices $\hat W_u $ and $\hat W_l$.
We have,  $  f_u =f_u (\theta_u, \varphi_u, \alpha_u)= \exp \left[  - (\theta_u^2 +\alpha_u^2)/(2 \rho_u^2) \right] /(2\pi)^{3/2} \rho_u^3    ,$  with    $\rho_{u}^2= {N_{u} \rho_{0}^2/3}$.
 The function $f_{u}$  does not depend on $\varphi_{u}$ and  is normalized as
  $  \int      f_u  du =1,$ where   $du =\theta_u d\theta_u  d\alpha_u  d\varphi_u$. Expression for $f_{l}$ is obtained by replacement $u\to l$.
 [One can easily show that the same functions $f_{u,l}$ can be used for disorder averaging of classical formula, Eq.\   \eqref{Pcl}.]
  In Fig.~\ref{fig-disorder} we present the results
  of calculations for averaged polarization.
  We see that  sharp resonances in polarization  broaden  with increasing the strength of magnetic disorder, $\rho^{2} = (N_{u} +N_{l})\rho_{0}^2/3$.

A single impurity with $S$ matrix, given by Eq.~\eqref{SM}  of the main text, placed at position $x_{0}$
is described by the  transfer matrix  \eqref{eq:W}.
Assuming the subsequent averaging over $\varphi$ we  may conveniently redefine $\varphi=\varphi+\pi/2$.
Having  $N_{u}$  impurities at the upper shoulder,   characterized by  transfer matrices
$\hat W_{1}$, \ldots $\hat W_{N_{u}}$, we
determine the transfer matrix of the whole upper shoulder as
\[\hat W_{u} =\hat W_{1}\hat W_{2} \ldots\hat W_{N_{u}} , \]
and similarly for $\hat W_{l}$.
In the weak impurity limit non-commutative property of $\hat W_{j}$ is relaxed and we obtain Eq.~\eqref{theta-alpha}.
We define 2$\times$2 matrices  $\kappa = \mbox{diag}[e^{-\lambda/2},e^{\lambda/2}]$,   $\Lambda_1 =  \mbox{diag}[e^{ikL/2+i\pi\phi},e^{-ikL/2+i\pi\phi}]$ and the matrix of transmission amplitudes
    \begin{equation*}
  \hat t  = 2\sinh \lambda\; \Lambda_1 \kappa\,\hat W_{u} \,\kappa \,
  ( 1 - \kappa\, \Lambda_1^{2}\hat W_{l}\,\kappa^{2}\hat W_{u}\, \kappa)^{-1}   \sigma_{3} \,.
\end{equation*}
The transmission coefficients are expressed via elements of $\hat t$ as follows:  $T_{\alpha\beta} (\epsilon) =   |t_{\alpha\beta}|^{2} $.
Straightforward calculation leads then to Eq.~\eqref{P-two-imp}.

Let us now calculate distribution functions for parameters of $\hat W_u$ and $\hat W_l.$
To this end, we enforce the conditions \eqref{theta-alpha} above
by  writing
\be
 \int \frac{ds_{1}\,ds_{2}\,ds_{3}}{(2\pi)^{3}}
e^{ is_{1}(\theta_u \cos \varphi_u - \rho_{0} \sum  \sin \eta_n \cos{(\varphi_n -2kL x_n}))
}   \times
e^{ is_{2}(\theta_u \sin \varphi_u - \rho_{0} \sum    \sin \eta_n  \sin{(\varphi_n -2kL x_n}))
+is_{3}   ( \alpha_u - \rho_{0}   \sum   \cos \eta_n )  } . \nonumber
\ee

Averaging over the orientation of impurities is given by $\prod _{n} {\sin \eta_{n}  d\eta_n\,d\varphi_n}/{4\pi}$. Performing this integration and then integrating over $s_{1,2,3}$ in weak scatterers' limit,  we obtain the above formulas for  $f_u$ and $f_l$.
The average polarization is given by $\langle P_{z}\rangle = \int P_z f_u f_l ~ du\, dl$. We raise the denominator of $P_{z}$ to the exponent, $\lambda(\theta_l ^2-\theta_u^2)/(4\lambda^2\! +\ldots ) =\lambda(\theta_l ^2-\theta_u^2) \int_{0}^{\infty} dz\, e^{-z(4\lambda^2\! +\ldots )}$ and perform integration over $du\, dl$. The remaining integration over $x = 2z ( \rho_u^2+ \rho_l^2)$ reads
\be
\langle P_{z}\rangle =
\lambda {\cal A} \int_{0}^{\infty} \tfrac{dx}{(1+x)^{5/2}} \exp\left[ -\tfrac{2x}{ \rho^2} \left( \lambda^{2} +
\tfrac{\pi^{2} \delta\phi^{2}}{1+x} \right) \right]
= \lambda  {\cal A} \;  F \left[  {\pi \delta\phi}/\lambda ,   \rho/{2\lambda} \right] \,,
\nonumber
 \ee
 where
 \be
     F \left[\Phi , z \right] =
  \tfrac{z^{2}}{2  \Phi ^2}\mbox{Re} \left[ 1- \sqrt{\tfrac \pi2 }  {e^{\frac{(1-i\Phi)^2}{2 z^2}}
   \left(\tfrac1z -i \tfrac z\Phi\right) \mbox{erfc}\left[\tfrac{1-i \Phi
   }{\sqrt{2} z}\right]}    \right] \,.
\label{Polarization_explicit}
\ee
Here  $\rho^{2}=  \rho_u^2+ \rho_l^2 $ and asymmetry parameter $ {\cal A} = ( \rho_u^2- \rho_l^2)/( \rho_u^2+ \rho_l^2)$.
The compact form \eqref{Polarization_explicit} was obtained by expanding general expressions at small $\lambda $, $\delta\phi$. Making substitution $\pi\, \delta\phi \to \frac12 \sin2\pi \phi$ in \eqref{Polarization_explicit}, we restore the expected periodicity of $\langle P_{z}\rangle$.  Thus obtained function  is shown in Fig.~\ref{fig-disorder}  of the main text.  It is a good approximation of $\langle P_{z}\rangle$ in the whole range of $\phi$.

\end{widetext}


\begin{thebibliography}{78}
\bibitem [{com({\natexlab{b}})}]{comment1}%
  \BibitemOpen
  \href@noop {} {} \bibinfo {note} {In Ref.\cite{my-cond}
  we, for simplicity, considered the model of impurity with zero forward
  scattering phase, $\alpha=0$}\BibitemShut {NoStop}%
\bibitem [{com({\natexlab{c}})}]{commentP}%
  \BibitemOpen
  \href@noop {} {}\bibinfo {note} {Actually, $P_z$ is
  defined as $( T_{ \uparrow \uparrow } + T_{ \uparrow \downarrow} - T_{
  \downarrow \uparrow}- T_{ \downarrow\downarrow} )/2 \mathcal T$ and coincides
  with Eq.\ \eqref{polarization} in the case $T_{\uparrow \downarrow}=T_{
  \downarrow\uparrow}$, see Eq.\ \eqref{|tab|^2}}\BibitemShut {NoStop}%
\bibitem [{com({\natexlab{d}})}]{comment2}%
  \BibitemOpen
  \href@noop {} {} \bibinfo {note} {Parameter $\lambda$ is
  connected with tunneling transparency $\gamma$ used in \cite{my-cond} as
  follows: $\gamma=\tanh (\lambda/2) $}\BibitemShut {NoStop}%
\end{thebibliography}
\end{document}